\documentclass[showpacs,twocolumn,aps,prx,floatfix,superscriptaddress]{revtex4-1}
\usepackage{amsmath,amssymb,graphicx,bm,color,mathrsfs,verbatim}
\pdfoutput=1

\newcommand{\be}{\begin{equation}}
\newcommand{\ee}{\end{equation}}

\begin{document}

\title{Long-time behavior of isolated periodically driven interacting lattice systems}
\author{Luca D'Alessio}
\affiliation{Department of Physics, The Pennsylvania State University, University Park, PA 16802, USA}
\affiliation{Department of Physics, Boston University, Boston, MA 02215, USA}
\author{Marcos Rigol}
\affiliation{Department of Physics, The Pennsylvania State University, University Park, PA 16802, USA}

\begin{abstract}
We study the dynamics of isolated interacting spin chains that are periodically driven by sudden 
quenches. Using full exact diagonalization of finite chains, we show that these systems exhibit three 
distinct regimes. For short driving periods, the Floquet Hamiltonian is well approximated by the 
time-averaged Hamiltonian, while for long periods the evolution operator exhibits properties of random 
matrices of a Circular Ensemble (CE). In-between, there is a crossover regime. Based on a finite-size 
scaling analysis and analytic arguments, we argue that, for thermodynamically large systems and 
non-vanishing driving periods, the evolution operator always exhibits properties of the CE of random matrices. 
Consequently, the Floquet Hamiltonian is a nonlocal Hamiltonian with multi-spin interaction terms 
and the driving leads to the equivalent of an infinite temperature state at long times. These 
results are connected to the breakdown of the Magnus expansion and are expected to hold beyond 
the specific lattice model considered.
\end{abstract}
\maketitle

\section{Introduction}

Periodically driven systems have a long history, one paradigmatic example 
is the Kicked-Rotor model of a particle moving on a ring subjected to time-periodic ``kicks'' 
\cite{kicked-rotor}. The behavior of such systems is very rich, e.g., they can display interesting
integrability-to-chaos transitions and dynamical Anderson localization \cite{chirikov_81,fishman_82,Reichl},
and counter intuitive effects such as dynamical stabilization \cite{Kapitza,Kapitza_math} both in 
classical and quantum mechanics. Moreover, it has been recently shown that periodic perturbations can be 
used as a flexible experimental knob \cite{AC-driven,mott-transition,Cavalieri} to realize synthetic 
matter, i.e., matter with specific engineered properties. Two outstanding examples in this direction 
are the opening of a gap in graphene by using light or carefully selected lattice phonons 
\cite{graphene0,graphene1,graphene2} and the realization of the so-called Floquet topological insulator 
in a material which, in absence of the driving, is topologically trivial \cite{natanel,natanel2}.

With a few exceptions \cite{prosen_1,prosen_2,prosen_3,prosen_4,joule_heating,luca-Magnus,sengupta,
Sandro1,Sandro2,Boris-Kai,Achille2,Dima}, recent studies mostly focus on driving simple (often 
single-particle) Hamiltonians. In some instances, it is known that the addition of interactions 
qualitatively changes the physics, e.g., when two quantum Kicked-Rotors are coupled the dynamical 
Anderson localization is suppressed and the (classical) diffusive behavior is recovered 
\cite{tworotators_th,tworotators_exp}. Furthermore, recent studies often rely on approximations 
that allow one to recast the effect of the drive in, e.g., an effective hopping 
\cite{effective-hopping-1,effective-hopping-2}. While these approximations can work well for 
non-interacting systems, their range of applicability for interacting quantum systems
is unclear. The validity of some of those approximations, such as the use of average Hamiltonians, 
has been a topic of interest because of its relevance to NMR experiments \cite{MAricq1,feldman}. 

The goal in this paper is to understand the long-time behavior of isolated interacting quantum systems 
periodically driven by sudden quenches. We focus on driving finite spin chains, 
for which the Hilbert space is finite. By considering chains of different lengths, we systematically study 
the role of finite size effects. Combining numerical results and analytic arguments, we argue that 
finite chains exhibit three distinct regimes, while in thermodynamically large systems only one regime 
survives and the system approaches the equivalent of an infinite temperature state at long times. In this 
regime, the Floquet Hamiltonian is a non-local Hamiltonian that contains multi-spin interaction terms. 
It is therefore qualitatively different from physically realizable static Hamiltonians, which generally 
contain only few-spin interaction terms. We argue that this regime is generic to interacting quantum 
systems. Thus, our study provides a paradigm to understand the long-time behavior of isolated periodically 
driven interacting quantum systems.

The exposition is organized as follows. In Sec.~\ref{theory}, we review the Floquet theorem and 
present general analytic arguments of relevance to our study. The numerical results for the specific model 
considered are presented in Sec.~\ref{sec:model}. Section \ref{sec:dis} is devoted to a discussion of 
far reaching implications of our study. A summary of our results is presented in Sec.~\ref{sec:summ}.


\section{Theoretical Analysis\label{theory}}

The Floquet theorem states that the evolution operator of any periodically driven system can be 
written as 
\be
\hat{U}(t)\equiv \hat{P}(t)e^{-i \hat{H}_{F}\,t/\hbar}, \label{Floquet-theorem}
\ee
where $\hat{P}(t+T)=\hat{P}(t)$ is a periodic unitary operator that reduces to the identity at 
stroboscopic times ($t_N=N\,T$, with $N$ integer), and $\hat{H}_{F}$ is the \textit{time-independent} 
Floquet Hamiltonian. It is clear that the factorization of the evolution operator is not unique and, 
therefore, there is some freedom in the definition of both the periodic operator $\hat{P}(t)$ and the 
Floquet Hamiltonian $\hat{H}_{F}$~\cite{Holthaus,marin_14}. At stroboscopic times, 
Eq.~\eqref{Floquet-theorem} simplifies as the periodic operator drops out. In particular, the exact 
evolution operator over a single driving cycle can be written as
\be   
\hat{U}_\text{cycle} = e^{-i \hat{H}_{F}\,T/\hbar}= \sum_{n}|\phi_n\rangle e^{-i\theta_n}\langle \phi_n|,
\label{eq:full-cycle}
\ee
where $|\phi_n\rangle$ and $e^{-i\theta_n}$ are the eigenstates and eigenvalues of $\hat{U}_\text{cycle}$.
From Eq.~\eqref{eq:full-cycle}, it follows that 
\be
\hat{H}_F\equiv\sum_{n}|\phi_n\rangle \varepsilon_n\langle \phi_n|, \label{eq:HF}
\ee
where $\varepsilon_n$ are the Floquet quasi-energies. The non-uniqueness of the factorization in
Eq.~\eqref{Floquet-theorem} translates into the fact that the Floquet quasi-energy $\varepsilon_n$ can 
be shifted by integers of $\hbar \omega$, where $\omega=2\pi/T$. We note, however, that this 
freedom does not affect neither the eigenstates $|\phi_n\rangle$ nor the eigenvalues 
$e^{-i\theta_n}=e^{-i \varepsilon_n T/\hbar}$ of $\hat{U}_\text{cycle}$, which are the focus of 
our study.

Equation~\eqref{eq:full-cycle} resembles the standard unitary evolution operator of a system described 
by a time-independent Hamiltonian $\hat{H}_F$. However, this simplicity is deceptive. 
For interacting systems, it is in general impossible to obtain the Floquet Hamiltonian in a closed form 
and one has to rely on approximations. A commonly used approximation scheme is the Magnus 
expansion~\cite{Magnus,Magnus-report,marin_14}, which is a series expansion in the driving period $T$. 
It allows one to compute the Floquet Hamiltonian as $\hat{H}_F=\sum_{n=0}^{\infty}\hat{H}_F^{(n)}$. 
The first two terms in the Magnus expansion are:
\be
\begin{split}
H_F^{(0)} &= \frac{1}{T}\int_0^T\mathrm{d}t\,\hat{H}(t) \\
H_F^{(1)} &= \frac{1}{2!\,Ti\hbar}\int_0^T\mathrm{d}t_1\int_0^{t_1}\mathrm{d}t_2\,
[\hat{H}(t_1),\hat{H}(t_2)], 
\label{eq:magnus_series}
\end{split}
\ee 
where $\hat{H}(t+T)=\hat{H}(t)$ is the time-periodic Hamiltonian of the system. 
The zeroth-order term, $\hat{H}_F^{(0)}$, is simply the time-averaged Hamiltonian, which we denote
as $\hat{H}_\text{ave}$ in what follows. Higher-order terms contain nested commutators of $\hat{H}(t)$ 
and multiple time-ordered integrals. The Magnus expansion is guaranteed to converge to the exact Floquet 
Hamiltonian for systems with bounded energy spectrum (such as the spin chain we consider here) and 
sufficiently short driving periods, i.e., $T\le T_c =2\pi/W$, where $W$ is the band-width of
$\hat{H}_(t)$ (see Ref.~\cite{Magnus-report} and references therein). For interacting systems, 
the energy bandwidth $W$ is extensive and therefore the Magnus expansion is guaranteed to converge 
only for $T \le T_c=\text{const}/V$, where $V$ is the volume of the system. We stress that this 
condition is only a sufficient one. It is unknown whether, for interacting systems in the 
thermodynamic limit, the radius of convergence of the Magnus expansion is finite. Our results 
suggest that this is in general not the case and that the Magnus expansion in thermodynamically 
large systems has a vanishing radius of convergence.

The Magnus expansion has several interesting properties~\cite{Magnus,Magnus-report}. First, being an 
expansion for the Floquet Hamiltonian it ensures that the evolution operator is unitary at any order 
in the expansion. This should be contrasted with the well-known Dyson series~\cite{Dyson}, which 
generates non-unitary evolution operators when truncated to any finite order. Second, noticing that 
the commutator of two local extensive operators is local and extensive, we see that the Floquet 
Hamiltonian is local and extensive in each order of the Magnus expansion. Therefore, when the 
Magnus expansion converges, the Floquet Hamiltonian is a local extensive Hamiltonian with few-body 
interactions. Hence, it exhibits the usual properties of experimentally realizable time-independent 
Hamiltonians~\cite{luca-Magnus}. However, when the Magnus expansion does not converge, the Floquet 
Hamiltonian can be qualitatively different from experimentally relevant static Hamiltonians, 
i.e., it can be ``unphysical" with nonlocal multi-body interactions.

To study properties of the Floquet Hamiltonian in regimes were the Magnus expansion 
might not converge, we follow the approach presented in Ref.~[\onlinecite{MAricq1}]. We introduce 
a generic bounded time-periodic local Hamiltonian $\hat{H}(\tau)$ defined on a finite-dimensional 
Hilbert space
\be
\hat{H}(\tau)=\hat{H}_\text{ave}+f(\tau)\hat{A},  \label{eq:Hamiltonian}
\ee
where $\tau\equiv t/T$ is the rescaled time, $f(\tau)$ is an arbitrary time-periodic function with 
zero average, $\hat{H}_\text{ave}$ is the time-averaged Hamiltonian, and $\hat{A}$ is a sum of local 
few-body operators. $|n(\tau)\rangle$ and $\epsilon_n(\tau)$ are the instantaneous eigenfunctions 
and eigenenergies of $\hat{H}(\tau)$:
\[
\hat{H}(\tau)|n(\tau)\rangle=\epsilon_n(\tau)|n(\tau)\rangle.
\]
The Schr\"odinger equation reads
\be
\partial_{\tau}\hat{U}(\tau)=T\,\frac{\hat{H}(\tau)}{i\hbar}\hat{U}(\tau),\label{schrodinger}
\ee
and is formally solved by
\be
\hat{U}(\tau)=\mathcal{T}\exp\left[T\,\int_{0}^{\tau}d\tau'\frac{\hat{H}(\tau')}{i\hbar}\right]
\equiv\exp\left[T\,\frac{\hat{H}_\text{eff}(\tau)}{i\hbar}\right],\label{eq:ansatz-2}
\ee
where $\mathcal{T}$ ensures that the exponential is time-ordered. 
The last equality defines an effective Hamiltonian $\hat{H}_\text{eff}(\tau)$. 
The instantaneous eigenstates and eigenvalues of the evolution operator $\hat{U}(\tau)$ are
\be
\hat{U}(\tau) |\phi_n(\tau)\rangle = e^{-i \theta_n(\tau)} |\phi_n(\tau)\rangle.\label{effective}
\ee
From Eqs.~\eqref{eq:full-cycle} and \eqref{eq:ansatz-2} one can see that 
$\hat{U}(\tau=1)=\hat{U}_\text{cycle}$ and, correspondingly,
$\hat{H}_\text{eff}(\tau=1)\equiv \hat{H}_F$. This justifies using the same notation for the 
eigenvectors and eigenvalues appearing in Eqs.~\eqref{eq:full-cycle} and \eqref{effective}.
Substituting Eq.~\eqref{eq:ansatz-2} into the  Schr\"odinger
equation~\eqref{schrodinger} leads to \cite{MAricq1} (see also Appendix~\ref{app_derivation})
\be
\begin{split}
\frac{i\hbar}{T}\,\left(1-e^{-i\left(\theta_n(\tau)-\theta_m(\tau)\right)}\right)&
\langle \phi_n(\tau)|\partial_{\tau}|\phi_m(\tau)\rangle= \\ 
&\langle \phi_n(\tau)|\hat{H}(\tau)|\phi_m(\tau)\rangle. \label{eq:fundamental}
\end{split}
\ee  
We note that this equation is exact since it corresponds to a rewriting of the Schr\"odinger equation. 

Remarkably, evaluating Eq.~\eqref{eq:fundamental} at $\tau=1$ allows one to connect the structure of 
the eigenstates of $\hat{U}_\text{cycle}$ (and of $\hat{H}_F$) to the statistics of the folded phases, 
i.e., the phases $\theta_n$ defined in $[-\pi,\pi)$. This connection can be made by analyzing under which 
circumstances the LHS and RHS are not identically zero. For example, when the eigenstates of $\hat{H}_F$ 
exhibit eigenstate thermalization \cite{ETH1_a,ETH1_b,ETH1_c,ETH2_a,ETH2_b,Lea-Marcos_a,Lea-Marcos_b}, 
the RHS of Eq.~\eqref{eq:fundamental} is generically non-zero. This occurs because $\hat{H}(\tau=1)$ is 
taken to be a sum of local few-body operators \cite{ETH1_a,ETH1_b,ETH1_c,ETH2_a,ETH2_b}. 
Therefore, the spectrum of $\hat{U}_\text{cycle}$ is expected to display repulsion, i.e., 
$e^{-i \left(\theta_n-\theta_m\right)}\ne 1$ or 
equivalently $\theta_n \ne \theta_m$. On the other hand, if $\hat{H}_F$ is noninteracting 
(or integrable) then the RHS of Eq.~\eqref{eq:fundamental} can be zero for a large fraction of 
pairs of states \cite{ETH2_a,ETH2_b} and there needs not be repulsion. In this case, the spectrum of 
$\hat{U}_\text{cycle}$ can exhibit highly degenerate phases separated by finite gaps. 

We should stress that, for $T>T_c=2\pi/W$, repulsion in the spectrum of $\hat{U}_\text{cycle}$ 
is incompatible with $\hat{H}_F=\hat{H}_\text{ave}$. This does not follow from 
Eq.~\eqref{eq:fundamental}. It becomes apparent because, for $\hat{H}_F=\hat{H}_\text{ave}$, 
$\theta_n=\theta_n^\text{ave}$, where $\theta_n^\text{ave}$ is obtained by folding 
$T\epsilon_n^\text{ave}/\hbar$ in $[-\pi,\pi)$, and $\epsilon_n^\text{ave}$ are the eigenvalues 
of $\hat{H}_\text{ave}$. Since energies $\epsilon_n^\text{ave}$ that are far apart are not expected 
to be correlated with each other, the (folded) phases $\theta_n$ would not exhibit repulsion. 
The same reasoning leads to the conclusion that, for $T>T_c$, phase repulsion is incompatible 
with $\hat{H}_F$ being an extensive operator with few-body interactions. Actually, repulsion in 
the phases of $\hat{U}_\text{cycle}$ for $T>T_c$ hints that $\hat{U}_\text{cycle}$ should exhibit 
properties of random matrices belonging to Circular Ensembles (CEs) 
\cite{COE_a,COE_b,COE_c,Mehta,Haake,RMT-Lea}. CEs are the equivalent of Gaussian Ensembles 
(GEs) for unitary matrices. In CEs, the phases display repulsion and the eigenstates are 
essentially random vectors.

Two remarks are in order. First, while it is believed that eigenstate thermalization and 
chaos (or level repulsion in the spectrum) come together in many-body interacting systems 
\cite{Lea-Marcos_a,Lea-Marcos_b,ETH1_a,ETH1_b,ETH1_c}, there is no proof that this is always the 
case. Our analysis based on Eq.~\eqref{eq:fundamental} does not constitute a proof that this is 
the case for $\hat{H}_F$. However, it is a step in the right direction. As long as $\hat{H}(\tau=1)$ 
is a sum on local operators and $T$ is small, Eq.~\eqref{eq:fundamental} allows us to make a connection 
between eigenstate thermalization and the presence of level repulsion in the spectrum. 
Second, phase repulsion in the Floquet spectrum is not unique to interacting systems. 
It can appear in chaotic single-particle driven systems (see for example Ref.~\cite{Haake})
for specific parameters of the driving protocol~\cite{Reichl}. 
For example in kicked systems the dynamics of the Floquet eigenvalues for varying kicked strength can be mapped to the dynamics of interacting quasi-particles, i.e. the ``Pechukas gas", and the phase repulsion in the Floquet spectrum can be inferred from the equilibrium properties of this fictitious gas~\cite{Haake}. 
What we expect to be different in driven interacting quantum systems is that phase 
repulsion will be robust to changes in the driving protocol.

We should also stress that the presence of level repulsion in the spectrum of $\hat{U}_\text{cycle}$, 
for non-vanishing values of $T$ in thermodynamically large systems, is an unbiased indicator of the 
breakdown of the Magnus expansion. This is because, as explained before, when the Magnus expansion 
converges $\hat{H}_F$ is a local extensive Hamiltonian with few-body interactions. This is 
incompatible with having phase repulsion for driving periods $T>T_c=\text{const}/V$, and 
$T_c$ vanishes with increasing system size.


\section{Numerical Results\label{sec:model}} 

We now turn to numerical simulations of a realistic many-body system. Specifically, we use full 
exact diagonalization to study a spin-$1/2$ chain with periodic boundary conditions and Hamiltonian
\begin{eqnarray}\label{eq:model}
&&\hat{H}(t)=\left[J+f(t)\,\delta J\right]\hat{H}_{nn}+
J^{\prime}\sum_{j}\sigma_{j}^{z}\sigma_{j+2}^{z}, \\
&&\hat{H}_{nn}=\sum_j \left[\sigma_{j}^{z}\sigma_{j+1}^{z}-
\frac{1}{2}\left(\sigma_{j}^{x}\sigma_{j+1}^{x}+
\sigma_{j}^{y}\sigma_{j+1}^{y}\right)\right], \nonumber
\end{eqnarray}
where $\sigma_j^\alpha$, $\alpha=x,y,z$, is the Pauli matrix for the $j^\text{th}$ spin, and
$f(t)$ is a periodic function with zero average
\be
f(t)\equiv\left\{ \begin{array}{ccc}
1 & \text{for} & N<\frac{t}{T}<N+\frac{1}{2}\\
-1 & \text{for} & N+\frac{1}{2}<\frac{t}{T}<N+1.\\
\end{array}\right. \label{eq:f_t}
\ee
The Hamiltonian $\hat{H}_{nn}$ in Eq.~\eqref{eq:model} is relevant to nuclear magnetic resonance experiments 
in solids \cite{NMR-exp}. The next-nearest neighbor coupling $J^{\prime}$ can be considered to be a 
phenomenological parameter that accounts for interactions that break the integrability of the 
spin-$1/2$ $XXZ$ chain. We restrict our calculations to the sector with total quasi-momentum $k=0$, 
magnetization $m_{z}=1/3$, and parity $p=+1$. For the four system sizes considered in our study, 
$L=15,\,18,\,21,$ and 24, the number of states in that sector are $D=111,\,561,\,2829,$ and $15581$, 
respectively.

\begin{figure}[!tb]
\includegraphics[width=0.9\columnwidth]{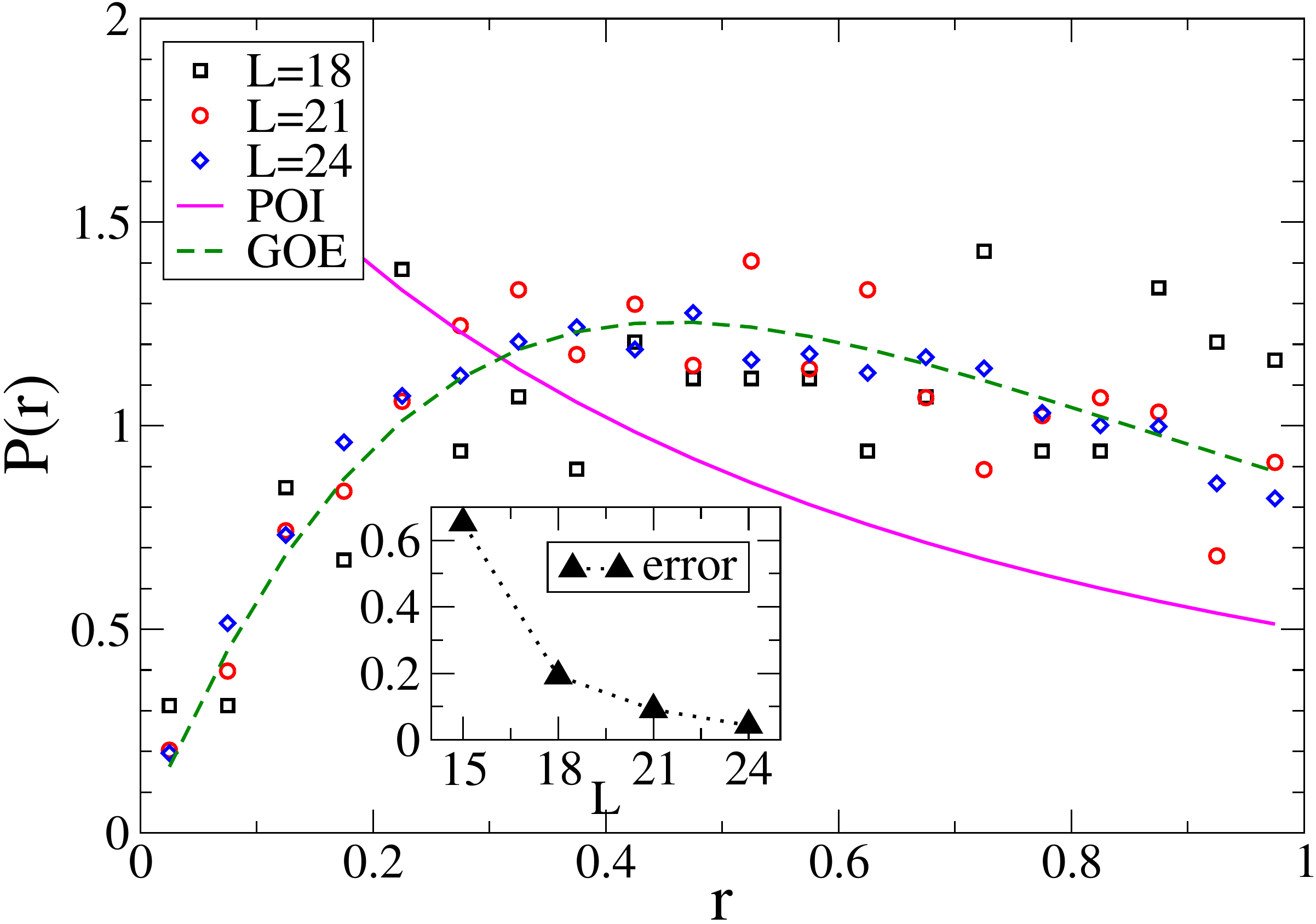}
\caption{(Color online) Probability distribution $P(r)$ for the ratio of two consecutive energy gaps 
in the spectrum of $\hat{H}_\text{ave}$ [see Eq.~\eqref{eq:r_def}] together with the GOE and POI 
predictions. (Inset) The error distance between $P(r)$ and the GOE prediction decreases quickly 
with increasing the system size.}
\label{fig:repulsion_hav}
\end{figure}

The Hamiltonian parameters are selected to be $J=1,\ \delta J=0.2,$ and $J^{\prime}=0.8$. They 
have been chosen to ensure that the average Hamiltonian is non-integrable and exhibits eigenstate 
thermalization and quantum chaos \cite{ETH1_a,ETH1_b,ETH1_c,ETH2_a,ETH2_b,Lea-Marcos_a,Lea-Marcos_b}. 
This is tested in Fig.~\ref{fig:repulsion_hav}, where we show the probability distribution $P(r)$, 
where $r$ is the ratio of two consecutive energy gaps~\cite{ratio_a,ratio_b,atas}, in the spectrum 
of $\hat{H}_\text{ave}$:
\be 
r=\frac{\min(s_n,s_{n+1})}{\max(s_n,s_{n+1})}\in[0,1],\quad 
s_n=\epsilon^{n+1}_\text{ave}-\epsilon^n_\text{ave}. \label{eq:r_def}
\ee
The distribution $P(r)$ is related to the level spacing distribution $P(s)$ in that they both measure 
the level repulsion in the spectrum, which is a standard quantum chaos indicator~\cite{Reichl}. However, 
the distribution $P(r)$ is simpler to compute since it does not require the unfolding of the spectrum, 
that is known to be non trivial~\cite{unfolding1,unfolding2}. In Fig.~\ref{fig:repulsion_hav}, 
we compare $P(r)$ to the predictions for the Gaussian Orthogonal Ensemble (GOE) and 
Poisson statistics (POI)~\cite{atas}:
\[
P_\text{GOE}(r)=\frac{27}{4}\,\frac{r+r^2}{(1+r+r^2)^{5/2}},\ \  P_\text{POI}(r)=\frac{2}{(1+r)^2}.
\]
We notice that $P(r)$ is well approximated by $P_\text{GOE}(r)$, and converges to the GOE prediction 
with increasing system size. This can be seen in the inset in which the error
\[
\text{error}=\int_0^1 dr |P(r)-P_\text{GOE}(r)|
\]
is shown to decrease as the system size increases. The presence of level repulsion is clearly seen as 
$r\rightarrow0$, where $P_\text{GOE}(r)$ vanishes while $P_\text{POI}(r)$ remains finite. The difference 
between those distributions is reflected in the average value of value of $r$:
\be
\langle r \rangle_\text{GOE}\approx0.535898,\quad \langle r \rangle_\text{POI}\approx 0.386294.
\ee 
As shown in Ref.~\cite{ratio_a,ratio_b}, $\langle r \rangle$ encodes information about level repulsion. 
It has been used as a sensitive and practical probe of the many-body localization 
transition~\cite{MBL1,MBL2,MBL3,MBL4,MBL5,MBL6}.  

Given our driving protocol [see Eq.~\eqref{eq:f_t}], the exact time-evolution operator over one cycle is
\be
\hat{U}_\text{cycle}=e^{-i \frac{T}{2\hbar} \hat{H}_{+}}\,e^{-i\frac{T}{2\hbar} \hat{H}_{-}}=
\sum_n |\phi_n\rangle e^{-i \theta_n} \langle\phi_n|,
\ee
where $\hat{H}_{\pm}$ corresponds to Eq.~\eqref{eq:model} for $f(t)=\pm1$. Using exact 
diagonalization, we obtain the eigenvectors $|\phi_n\rangle$ and the phases $\theta_n$, which 
we define in the interval $[-\pi,\pi)$. 

In what follows, we study how the properties of the time-evolution operator change as a function of the 
driving period $T$. We use indicators based on the phases (Sec.~\ref{phases}) and on the eigenvectors 
(Sec.~\ref{eigenvectors}) of $\hat{U}_\text{cycle}$. These two indicators suggest that 
$\hat{U}_\text{cycle}$ shares properties with matrices from the Circular Ensemble for nonvanishingly
small values of $T$ in thermodynamically large systems. This causes the system to approach, independently on the initial conditions, 
the equivalent of an infinite temperature state at long times (Sec.~\ref{heating}).


\subsection{Phase Repulsion\label{phases}}

We are first interested in understanding the properties of the phases $\theta_n$ as a function of 
the driving period $T$. We also want to find the relation between those phases and the phases
$\theta^\text{ave}_n$, which would be obtained if $\hat{H}_F=\hat{H}_\text{ave}$. To this end, 
we use the equivalent of $P(r)$ in our setup. We define $r$ as the ratio of two consecutive phase gaps: 
\be 
r=\frac{\min(\delta_n,\delta_{n+1})}{\max(\delta_n,\delta_{n+1})}\in[0,1],
\quad\delta_n=\theta_{n+1}-\theta_n. \label{phase_gap}
\ee
We stress that ``phase gaps" are obtained by first ordering the phases in $[-\pi,\pi)$ and then 
computing the difference between consecutive values.

\begin{figure}[!t]
\includegraphics[width=0.95\columnwidth]{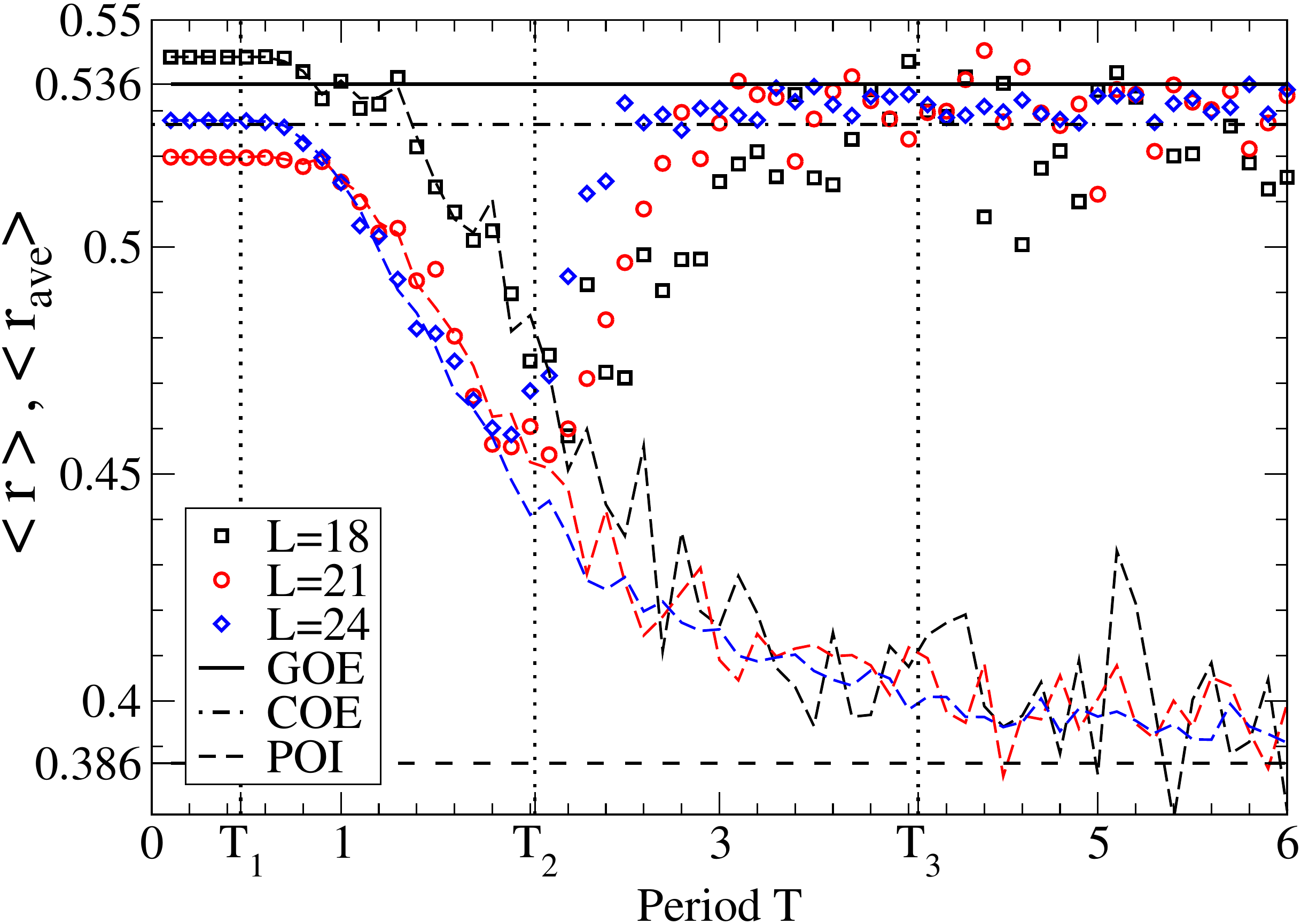}
\caption{(Color online) Average value of $r$ vs $T$. The values of $\langle r\rangle$ 
(symbols) and $\langle r_\text{ave} \rangle$ (dashed lines) are compared to the COE, GOE and POI 
predictions. The COE and GOE predictions are expected to coincide in the thermodynamic limit 
(see Appendix~\ref{app_repulsion}). Vertical dotted lines depict the periods $T_1=2\pi\hbar/W$, 
$T_2=\pi\hbar/\sigma$, and $T_3=2\pi\hbar/\sigma$ (see text).}
\label{fig:rav_nonint}
\end{figure}

In Fig.~\ref{fig:rav_nonint}, we show the average value of $r$ vs $T$ for the exact phases 
$\theta_n$ (indicated by $\langle r\rangle$) and the phases $\theta^\text{ave}_n$ (indicated by 
$\langle r_\text{ave}\rangle$). We compare them with the predictions of the Circular Orthogonal Ensemble 
(COE), the Gaussian Orthogonal Ensemble (GOE), and Poisson statistics (POI). The COE and GOE predictions 
in Fig.~\ref{fig:rav_nonint} have been computed for a finite size matrix. While the COE and GOE results 
obtained in those calculations are different, they are expected to coincide in the thermodynamic limit 
(see Appendix~B). 

In Fig.~\ref{fig:rav_nonint}, note the vertical dotted lines that mark 
$T_1=2\pi\hbar/W$, $T_2=\pi\hbar/\sigma$, and $T_3=2\pi\hbar/\sigma$, 
where $W=\epsilon_\text{ave}^\text{max}-\epsilon_\text{ave}^\text{min}$ is the band-width and 
$\sigma$ is the variance of spectrum of $\hat{H}_\text{ave}$ for $L=24$. The variance $\sigma$ is 
computed as the width of the best Gaussian fit to the density of states of $\hat{H}_\text{ave}$. 
We note that $W\propto L$ and $\sigma\propto\sqrt{L}$ scale differently with the system size, $L$.
The periods $T_1,T_2,$ and $T_3$ separate four regimes in which $\langle r\rangle$ and 
$\langle r_\text{ave}\rangle$ exhibit distinct behaviors. i) For $T<T_1$, $\langle r\rangle$ is 
independent of $T$, identical to $\langle r_\text{ave}\rangle$, and close to the 
GOE prediction. ii) For $T_1<T<T_2$, $\langle r\rangle$ and $\langle r_\text{ave}\rangle$ are very close 
to each other and decrease with increasing $T$. iii) For $T_2<T<T_3$, $\langle r\rangle$ increases towards 
the COE prediction while $\langle r_\text{ave}\rangle$ decreases towards the POI prediction. 
iv) For $T>T_3$, $\langle r\rangle$ is again independent of $T$ and close to the COE prediction, 
while $\langle r_\text{ave}\rangle$ is close and continues approaching the POI prediction.

\begin{figure}
\includegraphics[width=0.9\columnwidth]{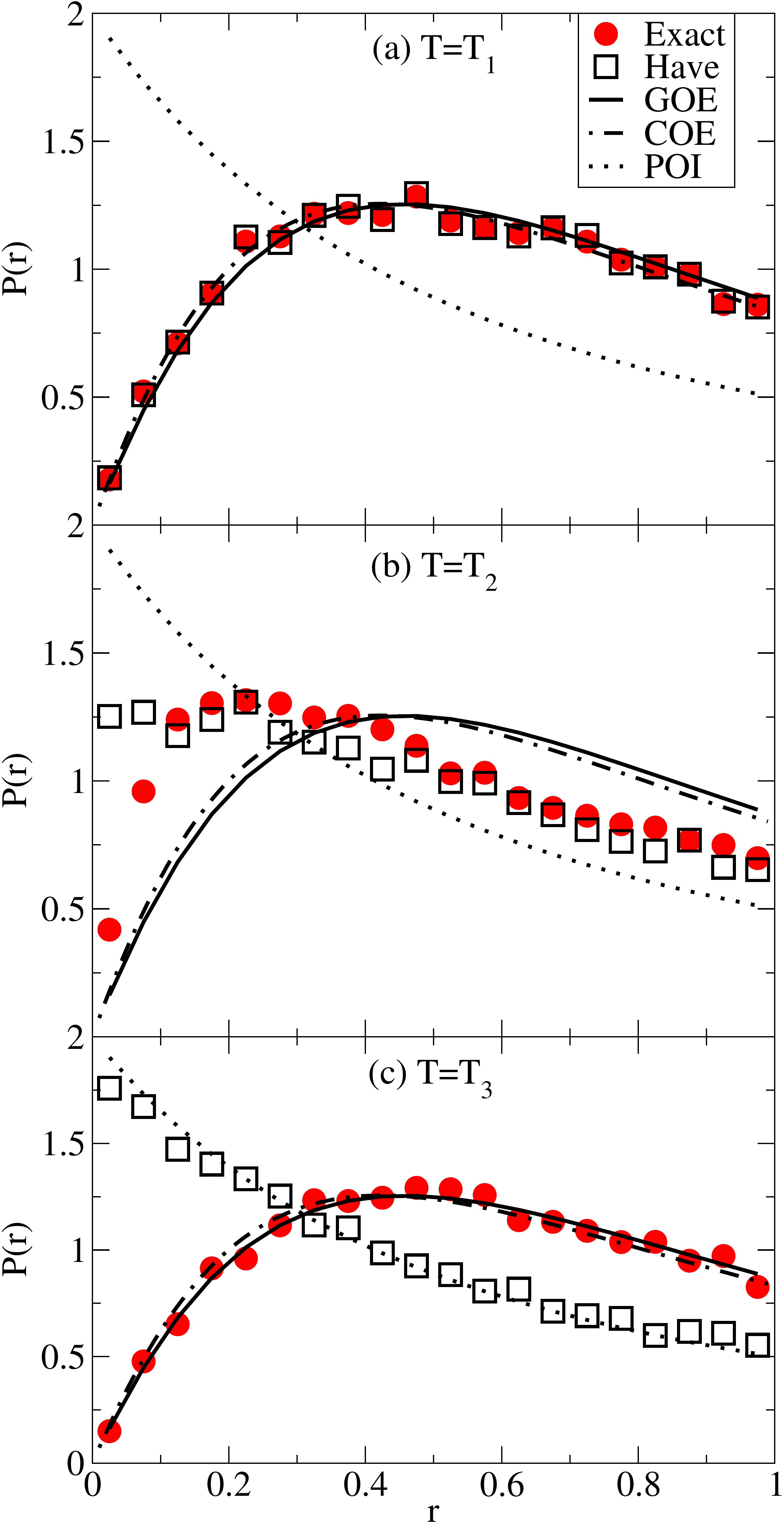}
\caption{(Color on-line) Distribution $P(r)$ for the periods $T=T_1,\,T_2,$ and $T_3$ shown 
in Fig.~\ref{fig:rav_nonint}. We compare the exact phases obtained from the exact diagonalization 
of $\hat{U}_\text{cycle}$ (indicated by red circles) and the phases one obtains if 
$\hat{H}_F=\hat{H}_\text{ave}$ (indicated by squares), with the GOE, COE and POI predictions (lines).}
\label{fig:Pr}
\end{figure}
 
In Fig.~\ref{fig:Pr} we show the full distribution $P(r)$ for both $\theta_n$ and $\theta_n^\text{ave}$
for the largest system size considered ($L=24$) for the periods $T_1,\,T_2,$ and $T_3$ (defined above).
We also compare them with the GOE, COE and POI predictions. These plots show that, for $T_1$ and $T_3$, 
the numerical results for the exact phases, $\theta_n$, are virtually indistinguishable 
from the GOE and COE predictions, respectively. On the other hand, the phases $\theta_n^\text{ave}$ are 
clearly described by Poisson statistics for $T_3$. For $T_2$, the results for both $\theta_n$ and 
$\theta_n^\text{ave}$ are in between GOE/COE and POI. Note that $T_1\propto L^{-1}$, $T_2\propto L^{-1/2}$, 
and $T_3\propto L^{-1/2}$ all collapse to zero in the thermodynamic limit.


\subsection{Eigenvectors' information entropy\label{eigenvectors}}

The properties of the eigenvectors of $\hat{U}_\text{cycle}$ provide further evidence of the 
different regimes seen in Fig.~\ref{fig:rav_nonint}. Specifically, we study how the eigenvectors of 
$\hat{U}_\text{cycle}$ differ from those of $\hat{H}_\text{ave}$ as $T$ is increased. We write the 
former in terms of the latter,
\[
|\phi_n\rangle=\sum_{m=1}^D c^n_m |m_\text{ave}\rangle,
\]
where $|m_\text{ave}\rangle$ are the eigenstates of $\hat{H}_\text{ave}$, 
$\hat{H}_\text{ave}|m_\text{ave}\rangle=\epsilon_m^\text{ave}|m_\text{ave}\rangle$,
and $D$ is the dimension of the Hilbert space. We then compute the (Shannon) information 
entropy \cite{Lea-Marcos_a,Lea-Marcos_b}
\be
S_n=-\sum_{m=1}^D |c^n_m|^2 \ln |c^n_m|^2,
\label{SS}
\ee
which measures the number of states $|m_\text{ave}\rangle$ that contribute to each $|\phi_n\rangle$. 
When $\hat{H}_F=\hat{H}_\text{ave}$, $S_n=0$. As $\hat{H}_F$ deviates from 
$\hat{H}_\text{ave}$, $S_n$ grows and is expected to saturate to the COE prediction 
$S_n\approx \ln(0.48 D)$, provided $\hat{U}_\text{cycle}$ exhibits properties of matrices from 
the Circular Ensemble.

\begin{figure}[!t]
\includegraphics[width=0.95\columnwidth]{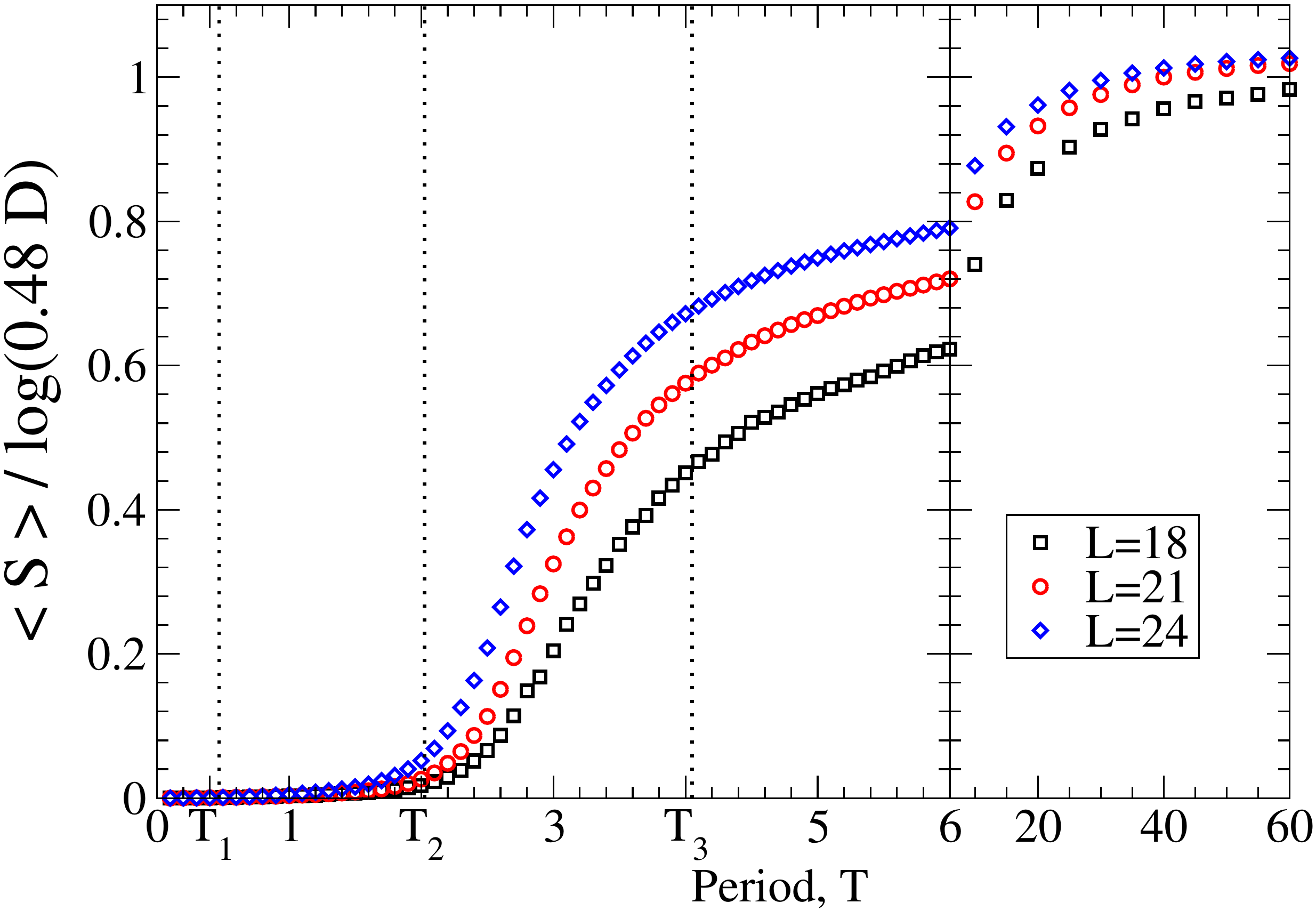}
\caption{(Color on-line) Average entropy of the eigenstates $\langle S \rangle$ vs the period $T$. 
$\langle S \rangle/\ln(0.48 D)$ crosses-over from zero to the COE prediction, which is reached 
earlier as $L$ increases. The variance $\langle (S_n-\langle S\rangle)^2\rangle^{1/2}$ is smaller 
than the size of the symbols used.}
\label{fig:Sdiag}
\end{figure}

In Fig.~\ref{fig:Sdiag}, we show the average entropy $\langle S\rangle=\left(\sum_{n=1}^D S_n \right)/D$ 
normalized by the COE prediction. Note that $\langle S\rangle$ grows monotonically with increasing 
$T$ and that the curves shift to the left (smaller periods) as the system size increases. The 
variance $\langle (S_n-\langle S\rangle)^2\rangle^{1/2}$ is smaller than 
the size of the symbols used in Fig.~\ref{fig:Sdiag}. This makes apparent that, in average, 
different eigenstates $|\phi_n\rangle$ delocalize the same way in the basis \{$|m_\text{ave}\rangle$\}. 
Contrary to $\langle r\rangle$ in Fig.~\ref{fig:rav_nonint}, $\langle S\rangle$ in Fig.~\ref{fig:Sdiag} 
is still changing for $T\ge T_3$. However, the COE prediction is reached for longer periods. 
This is not surprising since indicators based on the eigenvectors are known to suffer 
from stronger finite-size effects than level statistics indicators~\cite{Lea-Marcos_a,Lea-Marcos_b}. 


\subsection{Heating towards infinite temperature\label{heating}}

The fact that $\hat{U}_\text{cycle}$ shares properties with matrices from the COE has important 
consequences for how the system adsorbs energy. To show it, we compute the expectation value of 
$\hat{H}_\text{ave}$ at long times ($t=N T$ for $N\rightarrow\infty$). This can be done using 
the so-called diagonal ensemble \cite{ETH2_a,ETH2_b}
\begin{eqnarray}
\langle \hat{H}_\text{ave}\rangle(t=N T)&=&\langle \psi_0 | \left(\hat{U}^{\dagger}_\text{cycle}\right)^N 
\hat{H}_\text{ave}\,(\hat{U}^{}_\text{cycle})^N |\psi_0\rangle \nonumber \\ &\approx &\sum_n 
|\langle\psi_0|\phi_n\rangle|^2 \langle \phi_n|\hat{H}_\text{ave}|\phi_n\rangle,
\label{infT}
\end{eqnarray}
where $|\psi_0\rangle$ is the initial state. If $\hat{U}_\text{cycle}$ is COE like, all its 
eigenstates are close to random vectors and the eigenstate expectation values of few-body observables 
(such as $\hat{H}_\text{ave}$) are almost $n$-independent. As a result, the evaluation of 
Eq.~\eqref{infT} gives the same result one would obtain in a system at infinite temperature, 
and this occurs independently of the initial state selected. Instead, 
if $\hat{H}_F=\hat{H}_\text{ave}$, the system does not absorb energy under driving. 

\begin{figure}[!t]
\includegraphics[width=0.95\columnwidth]{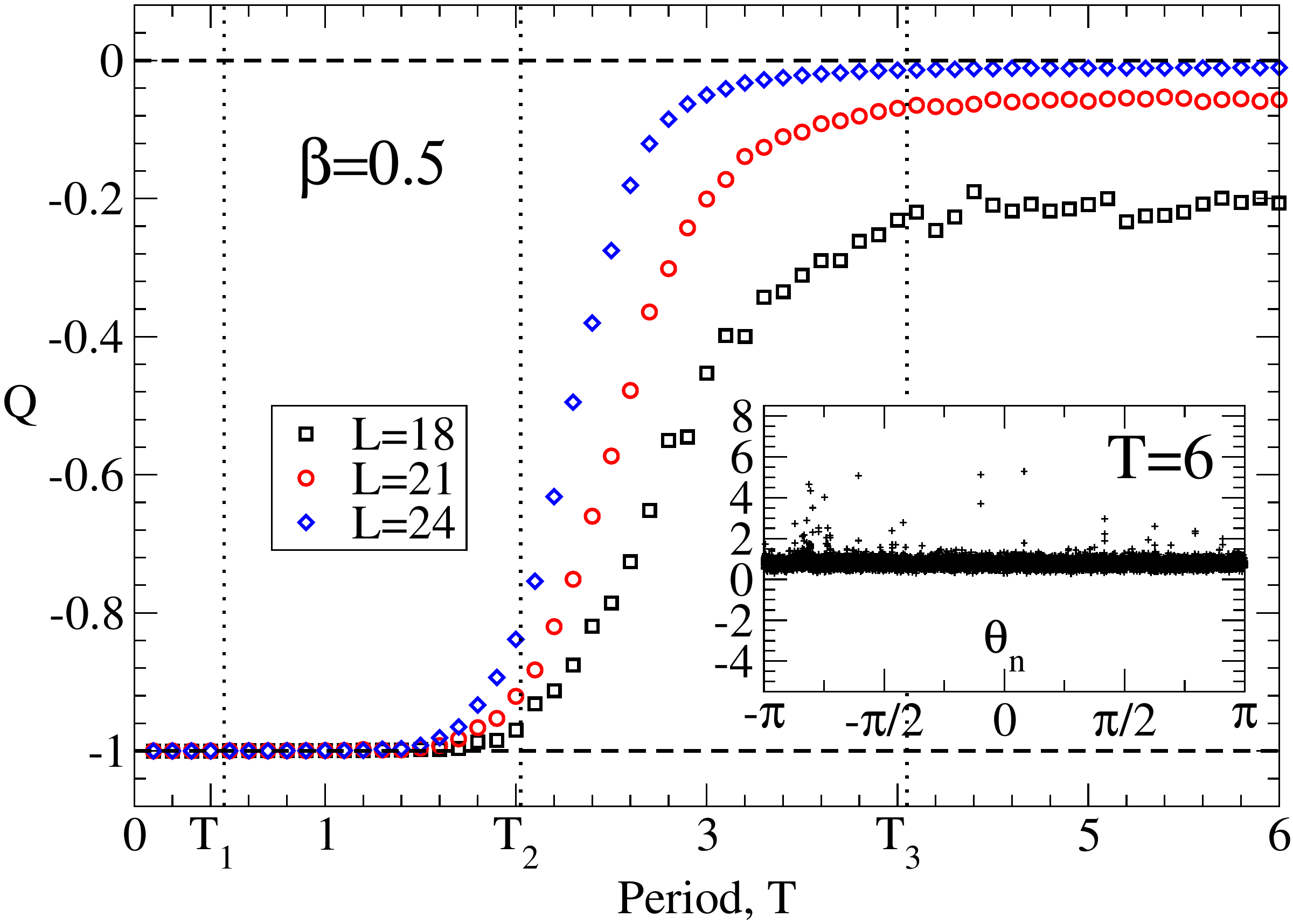}
\caption{(Color on-line) Absorbed energy. (Main) Absorbed energy at long (infinite in our calculation) 
times $Q$ [see Eq.~\eqref{eq:Qdef}] vs the driving period $T$. The initial state is a thermal state 
(with respect $\hat{H}_\text{ave}$) with $\beta=0.5$. (Inset) Expectation values 
$\langle\phi_n|\hat{H}_\text{ave}|\phi_n\rangle$ vs the exact phases $\theta_n$ 
for $L=24$ and $T=6$.}
\label{fig:EN}
\end{figure}

In Fig.~\ref{fig:EN}, we show the energy absorbed at long times $Q$ vs the period $T$. We have 
normalized $Q$ so that $-1$ corresponds to no-energy absorption and $0$ corresponds to the final 
energy being equal to that at infinite temperature:
\be
Q=\frac{\langle \hat{H}_\text{ave}\rangle(t\rightarrow\infty)-\langle \hat{H}_\text{ave}
\rangle_{\beta=0}}{\langle \hat{H}_\text{ave}\rangle_{\beta=0}-\langle \hat{H}_\text{ave}\rangle(t=0)}.
\label{eq:Qdef}
\ee
The initial state is taken to be in thermal equilibrium (with respect to $\hat{H}_\text{ave}$) with 
$\beta=0.5$. Namely, $|\psi_0\rangle\langle\psi_0|=Z^{-1}\sum_{m} e^{-\beta E_m^\text{ave}}
|m_\text{ave}\rangle\langle m_\text{ave}|$, where $Z$ is the partition function. Figure~\ref{fig:EN} 
shows that, with increasing $T$, $Q$ approaches zero as expected. In addition, the value of $T$ 
at which $Q$ saturates (close) to zero decreases with increasing system size.

\begin{figure}[!t]
\includegraphics[width=0.8\columnwidth]{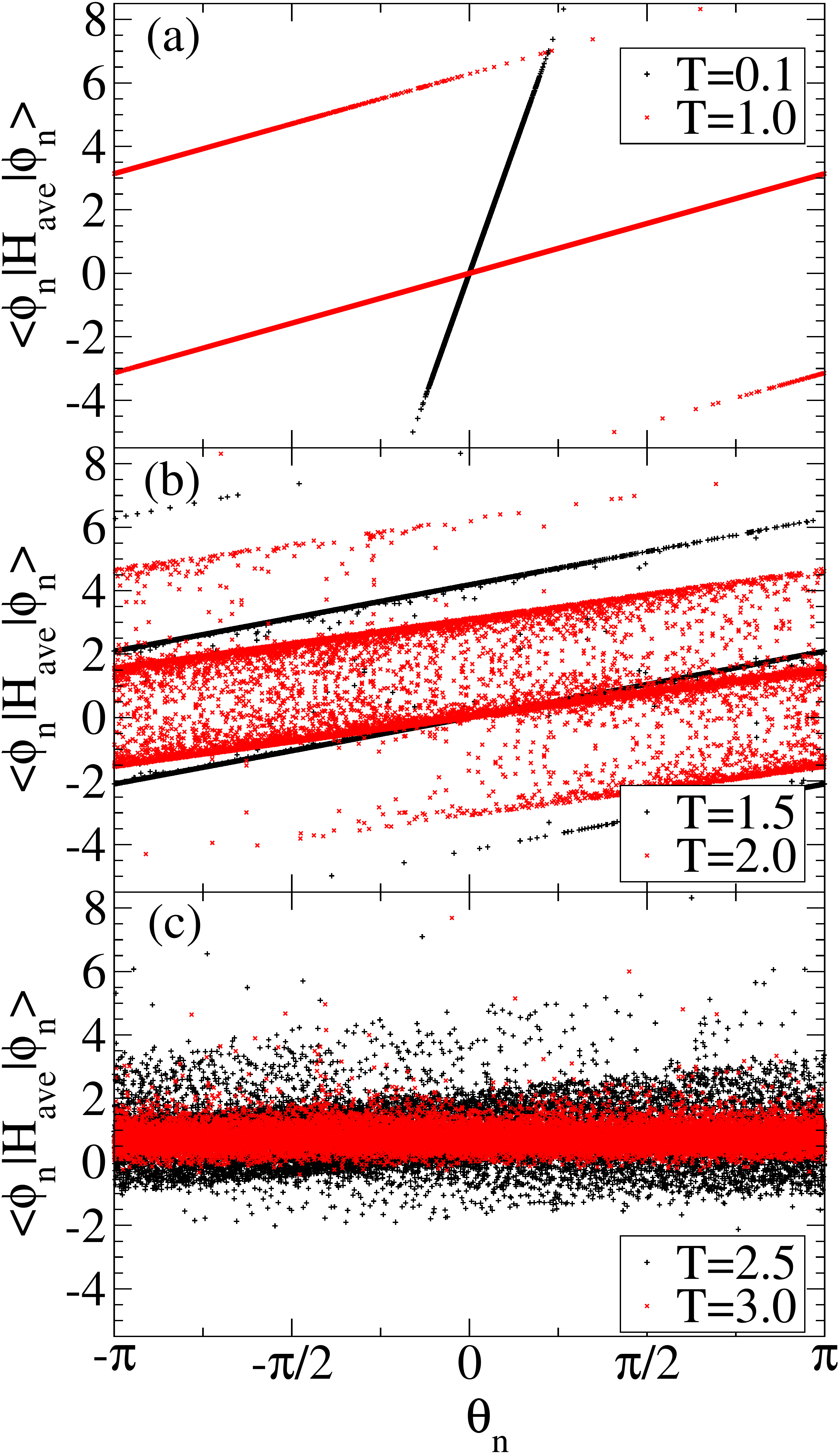}
\caption{(Color on-line) Expectation values $\langle \phi_n|\hat{H}_\text{ave}|\phi_n\rangle$ vs 
the exact phases $\theta_n$ folded in $[-\pi,\pi)$ for $L=24$ and different driving periods $T$.}
\label{fig:phases}
\end{figure}

As discussed above, the system reaching the equivalent of an infinite 
temperature state at long times can be traced back to the fact that, when $\hat{U}_\text{cycle}$ 
shares properties with matrices of the COE, the expectation value of few-body observables is 
nearly independent on the specific Floquet eigenstate. This can be seen in the inset 
in Fig.~\ref{fig:EN}, in which we plot the expectation values 
$\langle \phi_n|\hat{H}_\text{ave}|\phi_n\rangle$ vs $\theta_n$ for $L=24$ and $T=6$.

An understanding of how those eigenstate expectation values behave for different values of $T$ 
can be gained from Fig.~\ref{fig:phases}. There we report results for six different values 
of $T$. For short periods, the Floquet eigenstates, $|\phi_n\rangle$, are closely related 
to the eigenstates of $\hat{H}_\text{ave}$, and the simple relation 
\[
\theta_n=\frac{T}{\hbar}\,\langle \phi_n|\hat{H}_\text{ave}|\phi_n\rangle.
\]
holds [see Fig.~\ref{fig:phases}(a)]. We note that the phases $\theta_n$ might or might not span the 
entire ``Floquet Brillouin zone", i.e., $[-\pi,\pi)$, depending on the value of the period $T$. This 
can be also seen in Fig.~\ref{fig:phases}(a). For intermediate periods, there is a crossover 
regime in which the relation between $\theta_n$ and $\langle \phi_n|\hat{H}_\text{ave}|\phi_n\rangle$ 
is progressively lost with increasing $T$ [see Fig.~\ref{fig:phases}(b)]. For long periods $T$, the 
eigenstate expectation values $\langle \phi_n|\hat{H}_\text{ave}|\phi_n\rangle$ approach an 
$n$-independent value, i.e., they are not related to the values of the Floquet phases $\theta_n$ 
[see Fig.~\ref{fig:phases}(c)]. Increasing the period beyond the ones reported in Fig.~\ref{fig:phases} 
further decreases the eigenstate to eigenstate variation of 
$\langle \phi_n|\hat{H}_\text{ave}|\phi_n\rangle$, as one can realize by comparing 
the results in Fig.~\ref{fig:phases}(c) with those in the inset in Fig.~\ref{fig:EN}.


\section{Discussion\label{sec:dis}} 

Our numerical results for finite systems indicate that $\hat{H}_F=\hat{H}_\text{ave}$ for small 
values of $T$ ($T\lesssim T_1$ in our study), while $\hat{U}_\text{cycle}$ shares properties with matrices 
of the COE for large values of $T$ ($T\gtrsim T_3$ in our study). In addition, there is a crossover regime 
(for intermediate periods $T_1\lesssim T\lesssim T_3$) in which some of the indicators used smoothly 
interpolate between the values obtained in the previous two regimes. In the intermediate regime, a drive 
may lead to states in which the average energy in a cycle fluctuates between consecutive cycles or 
becomes stationary but not equal to that of a system at infinite temperature. This could potentially 
be seen in current experiments with periodically driven ultracold atomic systems. 

As we discussed in Sec.~\ref{theory}, when the Magnus expansion converges $\hat{H}_F$ is a local extensive 
Hamiltonian with few-body interactions, which is incompatible with the phases of the eigenvalues of 
$\hat{U}_\text{cycle}$ exhibiting level repulsion for non-vanishing values of $T$ in thermodynamically
large systems. Therefore, the fact that we see an onset of COE properties for $\hat{U}_\text{cycle}$
when $T\gtrsim T_3\propto 1/\sqrt{L}$ indicates that the radius of convergence of the Magnus expansion 
is vanishingly small in thermodynamically large systems. As a result, for non-vanishingly small values of
$T$, thermodynamically large systems reach the equivalent of an infinite temperature state at long driving 
times. Also, since the Floquet Hamiltonian $\hat{H}_F$ cannot be extensive with few-body interactions, 
one realizes that these systems offer a natural platform to investigate unique phenomena that may occur 
at intermediate times as a result of long-range many-body interactions in $\hat{H}_F$.

Remarkably, the Magnus expansion is guaranteed to converge only for $T\le T_1\propto 1/L$ 
\cite{marin_14,Magnus-report}, when the Floquet phases do not wrap around the 
``first Brillouin zone", i.e., $\theta_n\in[-\pi,\pi)$. Our 
results for $\langle r \rangle$, showing that $\langle r \rangle$ decreases approaching the POI
prediction when $T$ increases for $T\lesssim T_2$ [see Fig.~\ref{fig:rav_nonint}], suggest instead that 
the radius of convergence of the Magnus expansion is $\sim T_2\propto 1/\sqrt{L}$. This means that, for 
$T_1<T\lesssim T_2$, the Magnus expansion converges despite the fact that the Floquet phases wrap around the 
``first Brillouin zone". When the Magnus expansion converges one can ``unwrap" the Floquet phases $\theta_n$, 
which can be unambiguously defined in the ``extended Brillouin zone", i.e., $\theta_n\in(-\infty,\infty)$ 
[see Fig.~\ref{fig:phases}(a)], so that the the relation $\theta_n=T \varepsilon_n/\hbar$ holds and both 
$\theta_n$ and the Floquet quasi-energy $\varepsilon_n$ [see Eq.~\eqref{eq:HF}] are extensive. 
When the Magnus expansion fails~\cite{luca-Magnus}, it is not possible to 
unambiguously unwrap the Floquet phases, which are therefore only defined in the ``first Brillouin zone" 
[see Fig.~\ref{fig:phases}(c)]. Therefore the breaking of the Magnus expansion coincides with the transition 
from a description in the ``extended Brillouin zone" to one in the ``first Brillouin zone".

We expect that our results, in particular the presence of the two limiting behaviors described above for 
short and long driving periods, will be valid beyond the specific model considered. In fact, at sufficiently 
short driving periods, the Magnus expansion is guaranteed to converge for any system with a bounded spectrum 
and therefore $\hat{H}_F\approx \hat{H}_\text{ave}$. On the other hand, the same way that generic 
nonintegrable systems are expected to thermalize under unitary dynamics independently of the 
initial state selected \cite{ETH1_a,ETH1_b,ETH1_c,ETH2_a,ETH2_b}, periodically driven generic nonintegrable 
systems are expected to heat up towards the equivalent of an infinite temperature state independently of the 
initial conditions. This expectation can be traced back to the second law of thermodynamics. Actually, 
if the driving period is longer than all relaxation time-scales then the system relaxes and 
effectively starts each driving cycle from a stationary state. It can be proven rigorously 
that, during any dynamical process that starts from a stationary state, the properly defined 
entropy can only increase~\cite{diag_ent}. Hence, at long times, the system reaches a state 
of maximum entropy, i.e., the equivalent of an infinite temperature state.
This, in turn, requires $\hat{U}_\text{cycle}$ to have properties of the Circular Ensemble 
of random matrix. 

We note, however, that there are specific classes of models which escape this general 
expectation. In those systems, $\hat{H}_F$ is a well-behaved local Hamiltonian over a finite range of 
driving periods. Examples are: (i) driven Hamiltonians that commute at different times, 
$\left[\hat{H}(\tau),\hat{H}(\tau')\right]=0$, which lead to $\hat{H}_{F}=\hat{H}_\text{ave}$ independently 
of $T$, (ii) fictitious time-dependence that can be removed by a proper choice of the reference frame 
\cite{graphene2}, (iii) special cases in which multiple commutators of $\hat{H}(\tau)$ 
generate a finite algebra \cite{GAL,luca-Magnus}. In the presence of strong quenched disorder, interacting 
quantum systems can become non-ergodic. These systems, which are known as many-body localized (MBL), 
do not exhibit thermalization under unitary dynamics~\cite{MBL1,MBL2,MBL3,MBL4,MBL5,MBL6,khatami_rigol_12}. 
Currently, it is unknown whether MBL systems that are driven periodically in time by a global perturbation 
follow the general expectation discussed here (heat up indefinitely) or not~\cite{Iyer}. The case of a 
MBL systems driven periodically in time by local perturbations was studied in Ref.~\cite{Dima}, 
where they were shown not to heat up indefinitely. 

As discussed above, when the Magnus expansion does not converge, and $\hat{U}_\text{cycle}$ exhibits 
properties of random matrices of the Circular Ensemble, the system is expected to heat up 
towards the equivalent of an infinite temperature state at long times. We should stress, however, 
that the heating rate might be very small~\cite{MAricq1} and negligible on the time-scale of a 
particular experiment. If this is the case, then the experiment could be described using 
the Magnus expansion truncated after the first few orders~\cite{Boutis}. Moreover, at long times, the 
coupling between the system and the environment may become important and the system 
could approach a non-equilibrium steady state~\cite{Mitra,Assa} in which the energy absorbed 
from the driving is balanced by the energy dissipated into the environment~\cite{Holthaus}. 
Therefore the observation of the equivalent of an infinite temperature state requires a 
separation of time scales~\cite{marin_14}: $\tau_\text{heating}\ll \tau_\text{bath}$, where 
$\tau_\text{heating}^{-1}$ is the heating rate and $\tau_\text{bath}$ is the time-scale at 
which the coupling to the bath becomes important.

\section{Summary\label{sec:summ}}

In summary, we have studied the long-time behavior of an isolated interacting spin chain that is 
periodically driven by sudden quenches. For finite systems, we found three possible regimes. For short 
driving periods, $\hat{H}_{F}$ converges to the time-averaged Hamiltonian $\hat{H}_\text{ave}$, while 
for long periods the evolution operator $\hat{U}_\text{cycle}$ has properties of matrices of the COE
of random matrix theory. For intermediate periods, there is a cross-over regime. We have provided evidence 
that, for thermodynamically large systems, the only regime that occurs for non-vanishing driving periods 
is the one in which $\hat{U}_\text{cycle}$ exhibits properties of random matrices of the COE. This results 
in the system heating up to the equivalent of an infinite temperature state at long times. Furthermore, 
we argued that in this regime $\hat{H}_F$ cannot be an extensive Hamiltonian with few-body interactions.

Our findings for periodically driven systems are the equivalent of eigenstate thermalization for out 
of equilibrium many-body systems with time-independent Hamiltonians. While eigenstate thermalization 
ensures that thermalization occurs for generic interacting systems, the fact that the evolution 
operator in a periodically driven generic interacting system shares properties with matrices 
in the COE ensures that such systems thermalize to infinite temperature at long times.

\acknowledgements{We thank M. Bukov, I. Iadecola, A. Polkovnikov and L. F. Santos for 
stimulating discussions. This work was supported by the Office of Naval Research
and by the Army Research Office.}

\textit{Note added.} After posting our preprint on arXiv.org, preprints by Lazarides 
et al.~[\onlinecite{Achille2}] and Ponte et al.~[\onlinecite{Dima}] have also reported that 
periodically driven ergodic systems heat up leading to the equivalent of an infinite-temperature 
state, in agreement with our results.

\section*{Appendix}

\subsection{Derivation of Eqs.~\eqref{eq:fundamental}\label{app_derivation}} 

For completeness, we report the derivation of Eq.~\eqref{eq:fundamental}. Our presentation follows 
closely Ref.~\cite{MAricq1}. By plugging the ansatz~\eqref{eq:ansatz-2} in the  Schr\"odinger 
equation~\eqref{schrodinger} one obtains
\[
\partial_{\tau}\exp\left[T\,\frac{\hat{H}_\text{eff}(\tau)}{i\hbar}\right]=T\,\frac{\hat{H}(\tau)}{i\hbar}\,
\exp\left[T\,\frac{\hat{H}_\text{eff}(\tau)}{i\hbar}\right].\label{eq:step1}
\]
Using the mathematical identity
\[
\begin{split}
\partial_{\tau}\exp\left[\hat{B}(\tau)\right] &=\int_{0}^{1}d\alpha\,\,\left(\exp\left[\alpha\, 
\hat{B}(\tau)\right]\right. \\
&\times\left.\partial_{\tau}\hat{B}(\tau)\,\exp\left[(1-\alpha)\, \hat{B}(\tau)\right]\right),
\end{split}
\]
to compute the derivative of an exponential operator (note that this expression is required because 
$\hat{H}_\text{eff}(\tau)$ at different times may not commute) results in
\[
\begin{split}
\int_{0}^{1}d\alpha\,\,\left(\exp\left[\alpha\, T\,\frac{\hat{H}_\text{eff}(\tau)}{i\hbar}\right]\,
\partial_{\tau}\hat{H}_\text{eff}(\tau)\right. \\
\times\left.\exp\left[(1-\alpha)\, T\,\frac{\hat{H}_\text{eff}(\tau)} {i\hbar}\right]\right) 
=\hat{H}(\tau)\,\exp\left[T\,\frac{\hat{H}_\text{eff}(\tau)}{i\hbar}\right].\label{eq:step3}
\end{split}
\]
Multiplying both sides in the equation above by $\exp\left[-T\,\frac{\hat{H}_\text{eff}(\tau)}{i\hbar}\right]$ 
from the right one gets
\be
\begin{split}
\int_{0}^{1}d\alpha\,\,\left(\exp\left[\alpha\, T\,\frac{\hat{H}_\text{eff}(\tau)}{i\hbar}\right]\,
\partial_{\tau}\hat{H}_\text{eff}(\tau)\right. \\
\times\left.\exp\left[-\alpha\, T\,\frac{\hat{H}_\text{eff}(\tau)}{i\hbar}\right]\right)=\hat{H}(\tau).
\label{eq:step4}
\end{split}
\ee
We now close this equation between the exact (time-dependent) eigenstates of $H_\text{eff}(\tau)$ 
\be
\hat{H}_\text{eff}(\tau)|\phi_n(\tau)\rangle=\varepsilon_{n}(\tau)|\phi_n(\tau)\rangle, \label{eigen_Heff}
\ee
and perform the integration over $\alpha$ to obtain:
\be
\begin{split}
\frac{i\hbar}{T}\,\frac{e^{\frac{T}{i\hbar}\,\left(\varepsilon_n(\tau)-\varepsilon_m(\tau)\right)}-1}
{\varepsilon_n(\tau)-\varepsilon_m(\tau)} &\langle \phi_n(\tau)|\partial_{\tau}\hat{H}_\text{eff}(\tau)
|\phi_m(\tau)\rangle= \\ &\langle \phi_n(\tau)|\hat{H}(\tau)|\phi_m(\tau)\rangle,
\end{split}
\label{eq:result-H-1}
\ee
To obtain~\eqref{eq:fundamental}, we substitute in Eq.~\eqref{eq:result-H-1} the generalized 
Hellmann-Feynman theorem
\begin{eqnarray}
&& \langle \phi_n(\tau)|\partial_\tau \hat{H}_\text{eff}(\tau)|\phi_m(\tau)\rangle \nonumber \\
&& = \left(\varepsilon_{m}(\tau)-\varepsilon_{n}(\tau)\right) \langle \phi_n(\tau)|\partial_\tau 
|\phi_m(\tau)\rangle.
\end{eqnarray}

We note that, while Eq.~\eqref{eq:result-H-1} depends on the convention used in the definition 
of the Floquet quasi-energy in Eq.~\eqref{eigen_Heff}, the final equation~\eqref{eq:fundamental} 
can be written exclusively in terms of the phase factors 
$e^{-i \frac{T}{\hbar} \epsilon_n(\tau)}=e^{-i \theta_n(\tau)}$,
which are independent on the convention used in the definition of $\varepsilon_n(\tau)$. 


\subsection{Phase Repulsion in Circular Ensembles\label{app_repulsion}}

Here, we derive the probability distribution $P(r)$ in 
different ensembles. In Gaussian ensembles (GEs), $r$ is the ratio of consecutive energy gaps 
[see Eq.~\eqref{eq:r_def}] while in the Circular ensembles (CEs) $r$ is the ratio of consecutive phase 
gaps [see Eq.~\eqref{phase_gap}]. The distribution $P(r)$ 
has been shown to be a sensitive and practical probe \cite{ratio_a,ratio_b,atas} of the integrability-to-chaos 
transition \cite{Reichl}. For the GEs, the distribution $P(r)$ has been recently calculated \cite{atas} 
starting from the joint distribution of the eigenvalues of a $3\times3$ random matrix. We do the 
calculation for a $3\times3$ random matrix in the CEs. 

\begin{figure}[!t]
\includegraphics[width=0.45\columnwidth]{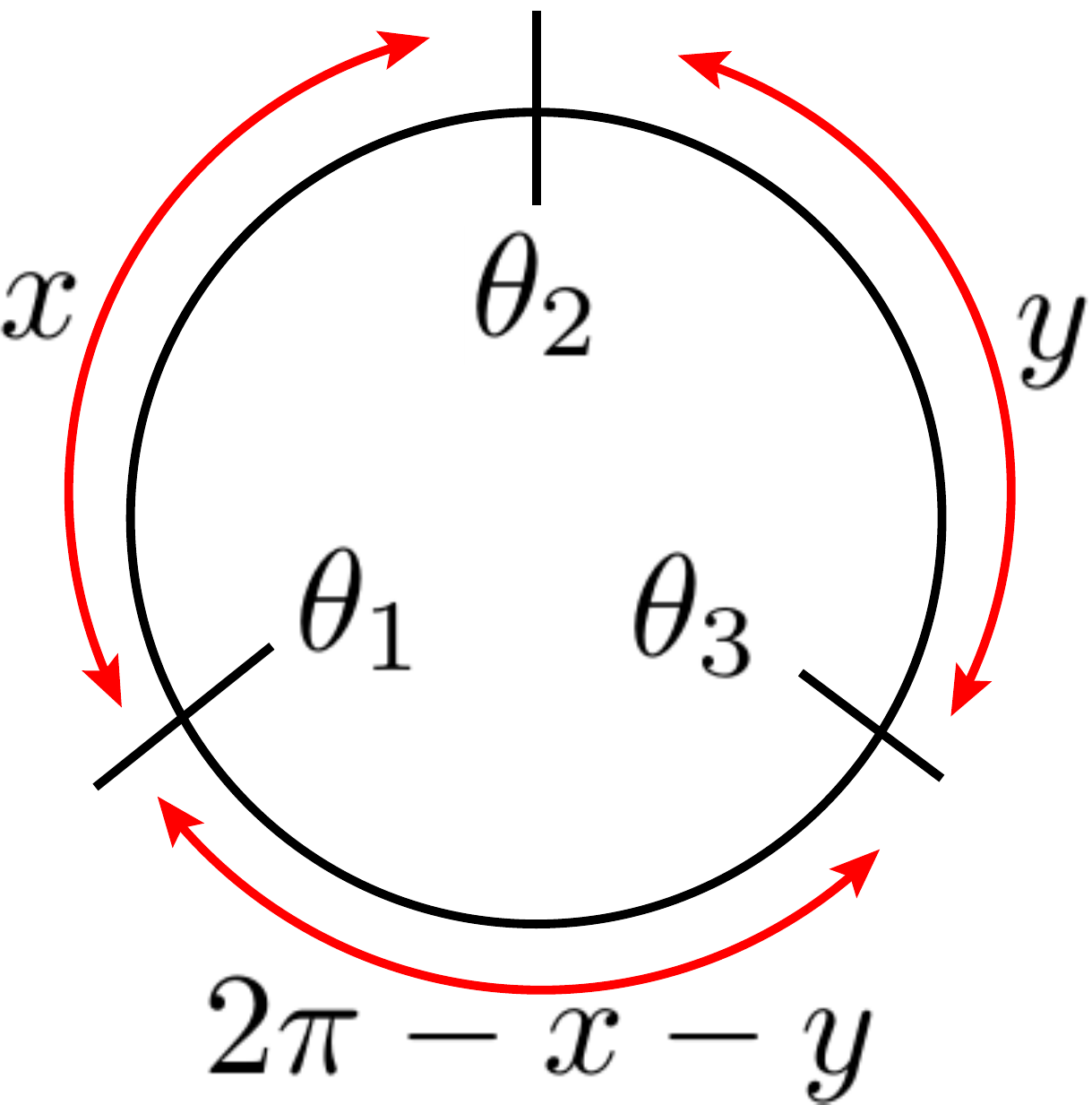}
\caption{(Color on-line) Transformation between $\theta_1,\theta_2,\theta_3$ and $x,\theta_2,z$.}
\label{fig:integration}
\end{figure}

The joint distribution for the phases in the Circular Orthogonal Ensemble (COE), the
Circular Unitary Ensemble (CUE), and Circular Symplectic Ensemble (CSE) 
are~\cite{COE_a,COE_b,COE_c,Mehta,Haake,RMT-Lea}:
\[
\begin{split}
\rho(\theta_{1},\cdots,\theta_{n})&=\frac{1}{Z_{n,\beta}}\prod_{1\leq k<j\leq n}|
e^{i\theta_{k}}-e^{i\theta_{j}}|^{\beta} \\
Z_{n,\beta}&=(2\pi)^{n}\frac{\Gamma\left(\frac{\beta n}{2}+1\right)}
{\left[\Gamma\left(\frac{\beta}{2}+1\right)\right]^{n}},
\end{split}
\]
and $\beta=1,2,4$ for COE, CUE and CSE, respectively. For a $3\times3$ matrix the expression simplifies to: 
\[
\begin{split}
\rho(\theta_{1},\theta_{2},\theta_{3})&=\frac{2^{3\beta}}{Z_{3,\beta}}\,\,
\left|\sin\left(\frac{\theta_{1}-\theta_{2}}{2}\right)\right|^{\beta} \\
&\times\left|\sin\left(\frac{\theta_{1}-\theta_{3}}{2}\right)\right|^{\beta}
\left|\sin\left(\frac{\theta_{2}-\theta_{3}}{2}\right)\right|^{\beta},
\end{split}
\]
where $\theta_i\in(0,2\pi)$ [this follows from the normalization $\iiint_0^{2\pi} d\theta_1d\theta_2d\theta_3 
\rho(\theta_1,\theta_2,\theta_3)=1$]. We choose one of the possible six ordering of the angles, 
$\theta_{1}\le\theta_{2}\le\theta_{3}$, define the variable of interest in one of the possible three ways 
$\delta\left(\tilde{r}-\frac{\theta_2-\theta_1}{\theta_3-\theta_2}\right)$, and perform the change of variables 
$x=\theta_{2}-\theta_{1},\theta_{2},y=\theta_{3}-\theta_{2}$ to obtain: 
\be
\begin{split}
P(\tilde{r})=\frac{2\pi}{3}\,\frac{6\,2^{3\beta}}{Z_{3,\beta}}\,\int_{0}^{2\pi}dx\int_{0}^{2\pi-x}dy\,
\delta\left(\tilde{r}-\frac{x}{y}\right) \\
\left(\sin\frac{x}{2}\,\sin\frac{y}{2}\,\sin\frac{x+y}{2}\right)^{\beta},
\label{eq:nice}
\end{split}
\ee
where the factor of $2\pi$ comes from the integration $\int_0^{2\pi} d\theta_2$, the factor of $6$ comes from 
the choice of the ordering, and the factor of $3$ comes from the choice of the observable. We also used the 
identity $\sin\frac{2\pi-x-y}{2}=\sin\frac{x+y}{2}$. The limits of integration in Eq.~\eqref{eq:nice} come from 
the fact that $0\le x+y\le 2\pi$ (see Fig.~\ref{fig:integration}).

\begin{figure}[!t]
\includegraphics[width=0.8\columnwidth]{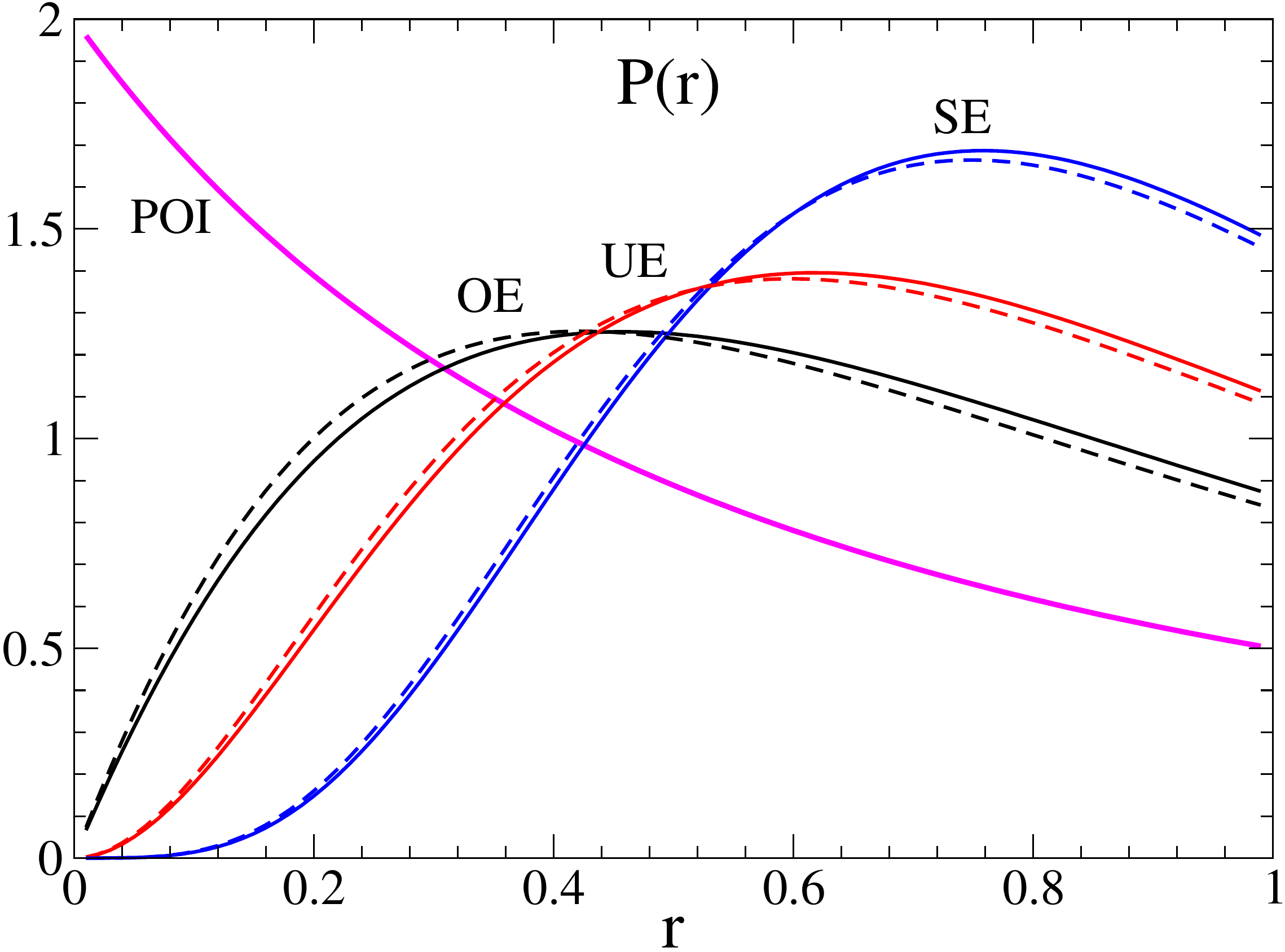}
\caption{(Color on-line) The distribution $P(r)$ in the Gaussian and Circular ensembles of random matrix theory
and for a sequence of Poisson distributed random numbers. The label POI refers to Poisson while the labels 
OE, UE, SE refer to Orthogonal, Unitary, and Symplectic Ensembles, respectively. Continues lines depict results 
for the Gaussian ensembles while dotted lines depict results for the Circular ensembles.}
\label{fig:distribution}
\end{figure}

In Eq.~\eqref{eq:nice}, $\tilde{r}\in(0,\infty)$ and is different from $r$ introduced in Eqs.~\eqref{eq:r_def} 
and \eqref{phase_gap}, which is defined in $[0,1]$. Luckily, the relation between $P(\tilde{r})$ and $P(r)$ is 
a simple one, $P(r)=2 P(\tilde{r})\Theta(1-\tilde{r})$ \cite{atas}. Using \textit{Mathematica}, we have 
performed the integration in Eq.~\eqref{eq:nice} exactly to obtain $P(r)$. For example, the distribution 
for the COE is:
\[
\begin{split}
P_\text{COE}(r)&=\frac{2}{3} \left(\frac{\sin \left(\frac{2 \pi  r}{r+1}\right)}{2 \pi r^2}+
\frac{1}{(r+1)^2}+\frac{\sin \left(\frac{2 \pi }{r+1}\right)}{2 \pi }\right. \\
&-\left.\frac{\cos \left(\frac{2 \pi}{r+1}\right)}{r+1}-
\frac{\cos \left(\frac{2 \pi  r}{r+1}\right)}{r (r+1)}\right).
\end{split}
\]
For brevity, we do not present the expressions for the CUE and CSE.

In Fig.~\ref{fig:distribution}, we plot the distributions $P_\text{COE}(r)$, $P_\text{CUE}(r)$, 
$P_\text{CSE}(r)$ and compare them with the corresponding distributions for the GEs 
[$P_\text{GOE}(r)$, $P_\text{GUE}(r)$, $P_\text{GSE}(r)$] and the distribution for a Poisson sequence 
of random numbers, $P_\text{POI}(r)$. The distributions for the CE and the corresponding GE are very 
close for $3\times3$ matrices and are expected to coincide in the thermodynamic limit. In fact, the phases 
in the Circular Ensembles (CEs) are expected to have the same local fluctuation properties as the eigenvalues 
of the corresponding Gaussian Ensembles (GEs) \cite{RMT-Lea}. In Table~\ref{table:r}, we 
report the average value of $r$ for all the distributions considered.

\begin{table}[!h]
\begin{tabular}{|c||c|c|c|}
\hline 
$\langle r\rangle$ & Orthogonal & Unitary & Symplectic  \tabularnewline
 \hline
Gaussian Ensembles & $0.535898$ & $0.602658$ & $0.676168$\tabularnewline
\hline 
Circular Ensembles & $0.526922$ & $0.596543$ & $0.671921$\tabularnewline
\hline 
\end{tabular}
\caption{(Color on-line) Average value of $r$ in the Gaussian and Circular ensembles 
of Random Matrix. For comparison the average value of $r$ for a sequence of Poisson distributed random 
numbers is 
$\langle r\rangle_\text{POI}=0.386294$.}
\label{table:r}
\end{table}

\end{document}